\documentclass[a4paper,12pt]{article}

\title{Eternally accelerating spacelike braneworld cosmologies}
\author{Riuji Mochizuki\thanks{e-mail address:rjmochi@tdc.ac.jp}
 \\  {\small  Laboratory of Physics, Tokyo Dental College, Chiba 261-8502, Japan}}

\date{\today}
\begin{document}

\maketitle
\abstract{We construct an eternally inflating spacelike brane world model.  If the space dimension of the brane is 
three (SM2) or six (SM5) for M theory or four (SD3) for superstring theory,  a time-dependent $n$-form field  would supply a constant energy density 
and cause exponentially expansion of the spacelike brane.  In these cases, the hyperbolic space 
perpendicular to the brane would keep its scale factor constant.   In the other cases, however, the extra space would vary in size.
\\
keywords: Spacelike braneworld, Accelerating cosmologies
  }
\section{Introduction}
Recent measurements of background cosmic microwaves \cite{infla2} have made it even clearer that our universe underwent an inflationary evolution  
during its infancy.  In addition, it has been shown that the present universe is still expanding, perhaps at a constant rate of acceleration (late-time inflation) 
\cite{cosmo}, although this rate is definitely
  much slower than initially.   The quantum effects of gravity may be essential in giving an account of this initial inflation.  Late-time inflation, however, 
 seems to be explained within the limits of general relativity, since the present universe is in a low-energy state.  If the cosmological constant is the only source
  of late-time inflation, it can not be zero but a very small positive value.  Although the cosmological constant is chosen arbitrarily within 
  the limits of general relativity, this becomes difficult if general relativity is regarded as a low-energy effective theory of M-theory/superstring theory.

There is a no-go theorem \cite{nogo} that 4-dimensional de Sitter spacetime could not realized by the ordinary compactification  methods 
available under M theory/superstring theory.  
One way to overcome this problem \cite{ngreview} would be to put S-branes, which are time-dependent spacelike branes, into the model 
\cite{sbrane}\cite{sbrane2}\cite{sbrane3}\cite{sbrane4}.  
Since D{\it p}-branes are objects which extend in {\it p} space dimensions and a time dimension, S{\it q}-branes extend in ($q+1$) space dimensions 
in ordinary notation.  It is known that accelerating S-brane solutions exist and that eternal inflation is possible if the universe is 
hyperbolic \cite{ohta}\cite{eternal}.  Higher order quantum corrections added to the action,  solutions including exponentially expanding braneworld and 
static extra
 space have been found numerically \cite{quan}\cite{quan2}.  Without higher order corrections, however, such solutions have not been found.

In this letter, our starting point is Einstein gravity coupled to a dilaton and an $n$-form field as a low-energy effective theory of M-theory/superstring theory.  
 Setting the time axis on the scale parameter of the hyperbolic space 
 perpendicular to the brane, we obtain analytical solutions 
 which include exponentially-expanding Robertson-Walker spacetime as the braneworld and hyperbolic space with a constant scale factor
  as the extra space, 
 provided the brane is SM2 or SM5 or SD3.  In the other cases, the braneworld also expands, but the extra space varies in size.  
 In the former cases, the dilaton coupling constant is zero and the $n$-form 
 field term works as a cosmological {\it constant}.  Some proposals for
  solving the cosmological constant problem 
 with the aid of 4-form fields of M-theory have been introduced \cite{4form1}\cite{4form2}.  The problem with these solutions, however, is the stabilization of the compact dimensions, 
 which is similar as suggested by other studies \cite{stab1}\cite{stab2}.   
 
 The hyperbolic space $H^{D-p-2}$ perpendicular to the exponentially-expanding 
 brane of our model has a constant scale factor.  
 Modding out $H^{D-p-2}$ by an appropriate freely acting discrete subgroup of the 
isometry group of $H^{D-p-2}$, a compact hyperbolic manifold is obtained.  A model in which the universe is the direct product of a Robertson-Walker 
spacetime and a compact hyperbolic manifold has been studied \cite{chm}\cite{chm2}\cite{chm3}.   These studies have pointed out the importance of compact
 hyperbolic manifolds for 
internal space, but have suggested no models for the origin of their structure.

\section{S-Brane solution}

We consider Einstein gravity coupled to a dilaton $\phi$ and an $n$-form field $F$ as a low-energy effective theory of M-theory/superstring theory, whose action
 $I$ is
\begin{equation}
I={1\over 16\pi G}\int d^Dx\sqrt{-g}\Big[ R-{1\over 2}(\partial \phi)^2-{1\over 2\cdot n!}e^{\alpha\phi}F^2\Big],\label{eq:action}
\end{equation}
where $\alpha$ is the dilaton coupling constant and the bare cosmological constant is assumed to be zero.  {\it D=11}
 for M-theory and {\it D=10} for superstring theories.

We assume the following metric form:
\begin{equation}
ds^2=e^{2u}\delta_{ij}dx^idx^j+e^{2v}\eta_{ab}dy^ady^b,\label{eq:setting}
\end{equation}
where
\[
i,j=1,\cdots,p+1,
\]
\[
a,b=p+2,\cdots,D,
\]
\[
\eta_{ab} = \{diag.(+ , \cdots , + , -)\}.
\] 
We use $x^i,\ i=1,\cdots,p+1$ as the coordinates on a spacelike brane (S{\it p}-brane).   A single S-brane solution  has been given \cite{single} 
and general orthogonally intersecting solutions have also been posited \cite{gen}, where the metric functions $u$ and $v$ 
 and fields $\phi$ and $F$ depend only on $y^D$.   We assume the metrics and the fields do not depend only on the timelike coordinate $y^D$
 but also on the other perpendicular coordinates $y^a,\ a=p+2,\cdots,D-1$, i.e.
 \[
u = u(y) \equiv u(y^{p+2},\cdots,y^D),
\]
\[
v = v(y) \equiv v(y^{p+2},\cdots,y^D),
\]
\[
\phi = \phi(y) \equiv \phi(y^{p+2},\cdots,y^D),
\]
\[
F = F(y) \equiv F(y^{p+2},\cdots,y^D).
\]
A D-brane solution which depends on all the extra space coordinates 
 has been suggested \cite{multi}.

The field strength for an electrically charged S{\it p}-brane is given by
\begin{equation}
F_{i_1\cdots i_{n-1}a}(y)=\epsilon_{i_1\cdots i_{n-1}}\partial_a E(y),
\end{equation}
where
\[
n = p+2.
\]
The magnetically charged case is given by
\begin{equation}
F^{a_1\cdots a_{n}}={1\over \sqrt{-g}}e^{-\alpha\phi}\epsilon^{a_1\cdots a_{n}b}\partial_b E(y),
\end{equation}
where
\[
n = D-p-2.
\]

The field equations are
\begin{equation}
-\partial^2 u-(\partial u)\big\{V(\partial u)+U(\partial v)\big\} = {U\over 2(U+V)}e^{\varepsilon\alpha\phi-2Vu}(\partial E)^2,\label{eq:eq1}
\end{equation}
\begin{eqnarray}
\eta_{ab}\Big[-\partial^2 v-(\partial v)\big\{U(\partial v)+V(\partial u)\big\}\Big] - U\partial_av\partial_bv-V\partial_au\partial_bu \nonumber \\
 -\big\{\partial_a\partial_b-\partial_av\partial_b-\partial_bv\partial_a\big\}\big\{Uv+Vu\big\} \nonumber \\
={1\over 2} e^{\varepsilon\alpha\phi-2Vu}\Big[\partial_aE\partial_bE-{V\over U+V} \eta_{ab}(\partial E)^2\Big]+{1\over 2}\partial_a\phi\partial_b\phi,
\end{eqnarray}
\begin{equation}
\partial_a\big\{e^{Vu+Uv} \eta^{ab} \partial_b\phi\big\}={\varepsilon\alpha \over 2}e^{\varepsilon\alpha\phi-Vu+Uv} (\partial E)^2,
\end{equation}
\begin{equation}
\partial_a\big\{e^{\varepsilon\alpha\phi-Vu+Uv} \eta^{ab} \partial_bE\big\}=0,\label{eq:eq2}
\end{equation}
and the Bianchi identity is 
\begin{equation}
\partial_{[a}F_{\cdots ]} = 0,\label{eq:bianchi}
\end{equation}
where $\eta^{ab}$ is the inverse matrix of $\eta_{ab}$ and 
\[
\partial^2 \equiv \eta^{ab}\partial_a\partial_b,
\]
\[
(\partial f)(\partial g) \equiv \eta^{ab}(\partial_a f)(\partial_b g)
\]
for arbitrary functions $f, g$.  $U$, $V$ and $\varepsilon$ are constants defined as 
\begin{equation}
U \equiv D-p-3,\ \ \ \ V \equiv p+1.\label{eq:UV}
\end{equation}
\begin{equation}
\varepsilon = +1 (-1)\ \ \ {\rm if}\ F\ {\rm is\ an\ electric\  (a\  magnetic)\  field}.
\end{equation}

To simplify the calculation, we assume 
\begin{equation}
Vu(y)+Uv(y)=0.\label{eq:assumption}
\end{equation}
Then, the above field equations and the Bianchi identity become
\begin{equation}
-\partial^2u={U\over 2(U+V)} e^{\varepsilon\alpha\phi-2Vu} (\partial E)^2,\label{eq:equ}
\end{equation}
\begin{eqnarray}
 &&-\partial^2v\eta_{ab} - U\partial_av \partial_bv-V\partial_au\partial_bu\nonumber\\
 &=& {1\over 2}\partial_a\phi\partial_b\phi + 
 {1\over 2} e^{\varepsilon\alpha\phi-2Vu} \Big[\partial_aE\partial_bE-{V\over U+V}\eta_{ab}(\partial E)^2\Big],\label{eq:eqv}
\end{eqnarray}
\begin{equation}
\partial^2\phi={\varepsilon\alpha\over 2} e^{\varepsilon\alpha\phi-2Vu}(\partial E)^2,\label{eq:eqphi}
\end{equation}
\begin{equation}
\partial_a\big\{e^{\varepsilon\alpha\phi-2Vu}\eta^{ab}\partial_bE\big\}.\label{eq:eqE}
\end{equation}
The Bianchi identity (\ref{eq:bianchi}) is trivially satisfied in the electric case, and so is the field equation (\ref{eq:eq2}) in the magnetic case.

We write candidates which satisfy the assumption (\ref{eq:assumption}) for the solution of the above equations (\ref{eq:equ}) $\sim$ 
(\ref{eq:eqE}) as
\begin{equation}
E(y)=\sqrt{{2\over W}}iH(y),\label{eq:solE}
\end{equation}
\begin{equation}
u(y)={U\over W(U+V)}\ln H(y),\label{eq:solu}
\end{equation}
\begin{equation}
v(y)=-{V\over W(U+V)}\ln H(y),\label{eq:solv}
\end{equation}
\begin{equation}
\phi(y) =-{\alpha\over W}\ln H(y),\label{eq:solphi}
\end{equation}
where
\begin{equation}
W={\alpha^2\over 2}+{UV\over U+V}.\label{eq:W}
\end{equation}
These become the right solutions of the field equations (\ref{eq:equ}) $\sim$ (\ref{eq:eqE})  if 
\begin{equation}
H(y) \equiv {1\over h(y)},\ \ \ \ \ \partial^2h(y)=0.\label{eq:h}
\end{equation}

\section{Eternally inflating brane}
Now consider a metric which depends only on the scale parameter $r$ of the entire, or part, of the  spacetime perpendicular to the brane:
\begin{equation}
r \equiv \sqrt{-\eta_{\tilde a\tilde b}y^{\tilde a}y^{\tilde b}},\ \ \ \ -\eta_{\tilde a\tilde b}y^{\tilde a}y^{\tilde b}>0,
\end{equation}
\[
\tilde a, \tilde b = D-m,\cdots,D,\ \ \ \ \ 2\leq m\leq D-p-2.
\]
Note that $r$ is a timelike coordinate.  To satisfy (\ref{eq:h}),
\begin{equation}
H(r) \sim r^{m-1}.\label{eq:HH}
\end{equation}
Then, the metric of this spacetime is 
\begin{equation}
ds^2_\bot = -e^{2v}[dr^2 - r^2\big(d\theta^2 + \sinh^2\theta(d\Omega_{m-1})^2\big) - \delta_{\hat a\hat b}dy^{\hat a}dy^{\hat b}], \label{eq:metric}
\end{equation}
\[
\hat a, \hat b = p+2,\cdots ,D-m-1,
\]
where $d\Omega_{m-1}$ is the metric on a $(m-1)$-dimensional unit spherical surface $S^{m-1}$ and $r^2\big(d\theta^2 + \sinh^2\theta(d\Omega_{m-1})^2\big)$
 is the metric of $m$-dimensional hyperbolic space $H^{m}$.

We will look for an exponentially inflating brane solution.  That is, $d$-dimensional Robertson-Walker spacetime  
\begin{equation}
ds_d^2 = -dt^2 + a^2(t)dx^2
\end{equation}
is constructed from an S-brane and the {\it time} perpendicular to the S-brane.  The scale factor $a(t)$ is assumed to be an increasing exponential of
 cosmic time $t$:
\begin{equation}
u = kt,\label{eq:k}
\end{equation}
where $k$ is a positive constant.

Let us define cosmic time $t$ as
\begin{equation}
dt^2 \equiv e^{2v(r)}dr^2.\label{eq:t}
\end{equation}
Solving (\ref{eq:k}) and (\ref{eq:t}), $H$ and $t$ are given by
\begin{equation}
H = \Big({Vkr\over U}\Big)^{W(U+V)/V},\label{eq:H}
\end{equation}
\begin{equation}
t = {U\over kV}\ln \Big({Vkr\over U}\Big).
\end{equation}

Taking into account $m\leq D-p-2$, the exponent in (\ref{eq:H}) should satisfy
\begin{equation}
{W(U+V)\over V} = m-1 \leq D-p-3.\label{eq:cond}
\end{equation}
Calling (\ref{eq:UV}) and (\ref{eq:W}), (\ref{eq:cond}) is satisfied if the dilaton coupling $\alpha=0$ and 
$m-1=D-p-3$.  $\alpha=0$ is the case in 11-dimensional supergravity and 10-dimensional superstring with a 5-form field.  The former has a 4-form field, which 
introduces SM2-brane in the electric case or SM5-brane in the magnetic case.  The latter includes SD3-brane in both cases.  The SM2-brane solution, in particular, gives 
an eternally accelerating 3-dimensional space.  

 Although our setting may seem to be similar to previous studies\cite{ngreview}\cite{sbrane4}\cite{quan}\cite{quan2}, our solution is quite novel.  
 If we start with metric such as
 \begin{equation}
 ds^2 = -e^{2u_0(t)}dt^2 + e^{2u(t)}\delta_{ij}dx^idx^j + e^{2v(t)}d\Sigma
 \end{equation}
 in place of (\ref{eq:setting}), our solution is not obtained as a general solution, that is, it becomes a singular solution of the field equations.

We expect the non-zero cosmological constant to appear when $m-1=D-p-3$ and $\alpha=0$, since the brane expands exponentially.  Let us confirm this.  
When the dilaton coupling $\alpha=0$, the dilaton field vanishes and the action (\ref{eq:action}) becomes
\begin{equation}
I={1\over 16\pi G}\int d^Dx\sqrt{-g}\Big[ R-{1\over 2\cdot n!}F^2\Big],
\end{equation}
where
\begin{equation}
F^2 = e^{-2Vu-2v}n! (\partial E)^2.
\end{equation}
With the aid of (\ref{eq:solE}), (\ref{eq:solu}), (\ref{eq:solv}) and (\ref{eq:set}), the $n$-form term becomes
\begin{equation}
{1\over 2n!}F^2 = {(m-1)^2\over W}\Big({Vk\over U}\Big)^2 H^{2({m-1\over U}-1)}.
\end{equation}
Hence, it works as a cosmological constant at a classical level.

Next, we consider the extra space.  When $m-1=D-p-3$ and $\alpha=0$,  the metric function of the hyperbolic space is
\begin{equation}
e^{2v} = \Big( {Vkr\over U}\Big)^{-2}.
\end{equation}
Cancelling $r$-dependence, the metric (\ref{eq:metric}) becomes
\begin{equation}
ds^2_{\bot} = -dt^2 + {U^2\over V^2k^2}\big(d\theta^2 + \sinh^2\theta(d\Omega_{D-p-3})^2\big).
\end{equation}
Thus, we see that the scale of the hyperbolic space is independent of time.

On the other hand, if $m-1 < D-p-3$ for superstring theory, setting 
\begin{equation}
H=\Big({Vkr\over U}\Big)^{m-1},\label{eq:set}
\end{equation}
the scale factor of the brane becomes
\begin{equation}
e^{u(t)} = \Big[\Big( 1 - {(m-1)V\over W(U+V)}\Big){Vkt\over U}\Big]^{{(m-1)U\over W(U+V)-(m-1)V}}.\label{eq:less}
\end{equation}
With the aid of (10) and (21), 
\[
1 - {(m-1)V\over W(U+V)}>0,
\]
\[
{(m-1)U\over W(U+V)-(m-1)V}>0,
\]
so the brane also expands in this case.  Moreover, when
\begin{equation}
m> 3,
\end{equation}
the scale factor of the brane $e^{u(t)}$ satisfies
\[
{d^2\over dt^2}e^{u(t)}>0,
\]
and the brane undergoes accerated expansion.   In this situation, however, the extra space varies in size.

To summarize, when $D=11$, $d=3$ and $d=6$ for M-theory and $D=10$, $d=4$ for superstring theory, the metric
\begin{equation}
ds^2 = -dt^2 + e^{2kt}\sum_{i,j=1}^d\delta_{ij}dx^idx^j + {U^2\over V^2k^2}\big(d\theta^2 + \sinh^2\theta(d\Omega_{D-d-2})^2\big)
\end{equation}
is the solution of the basic equations (\ref{eq:eq1})$\sim$(\ref{eq:bianchi}) and the $n$-form field strength term is constant and works as 
the cosmological constant at a classical level.   SM2-brane solution has a phenomenological significance, where {\it our} braneworld is 
(3+1)-dimensional Robertson-Walker spacetime and its extra space is hyperbolic and keeps its size constant.

\end{document}